\documentclass[%
preprint,
 amsmath,amssymb,
 aps,
]{revtex4-1}
\usepackage{graphicx}
\usepackage{dcolumn}
\usepackage{bm}
\begin{document}
\preprint{WM/CCR-V4}

\title{How to measure the canonical commutation relation $\mathbf{[\hat{x},\hat{p}]=i\hbar}$ in quantum mechanics with weak measurement?}
\author{Yiming Pan}
\email{yimingpan@mail.tau.ac.il}
\affiliation{Department of Electrical Engineering Physical Electronics,\\
Tel Aviv University, Ramat Aviv 69978, ISRAEL}

\date{\today}
\begin{abstract}
The quantum weak value draws many attentions recently from theoretical curiosity to experimental applications. Now we design an unusual weak measuring procedure as the pre-selection, mid-selection and post-selection to study the correlation function of two weak values, which we called the weak correlation function. In this paper, we proposed an weak measurement experiment to measure the canonical commutator $[\hat{x},\hat{p}]=i\hbar$ in quantum mechanics. Furthurmore, we found the intriguing equivalence between the canonical commutation relation and Riemann hypothesis, and then obtained the weak value of nontrivial Riemann zeros. Finally, as an nontrivial example of weak correlations, we also passed successfully a testing on the (anti-)commutators of Pauli operators, which followed the experimental setup of the landmark paper of Aharonov, et al. in 1988.  Our proposed experiments could hopefully test the fundamental canonical relationship in quantum worlds and trigger more testing experiments on weak correlations.
%
\end{abstract}
\maketitle


\section{Measuring the canonical commutator $[\hat{x},\hat{p}]=i\hbar$}
The weak measurement idea was first introduced by Y.Aharonov et al in 1988\cite{Yakir1988}. It was used to show how pre- and post-selection could weakly measure the outcome of operators without collapsed the system. Here, we follow the same idea to measure the fundamental canonical commutation relation (CCR). In quantum mechanics, the CCR is the fundamental postulated relation between canonical conjugate quantities, for instance, the position operator $\hat{x}$ and its conjugated momentum operator $\hat{p}$. Historically, the CCR served as a postulated of the quantum theory, which is attributed to Max Born (1925) and resulted in the Heisenberg uncertainty principle. The uncertainty relation implied that no measurement can be simultaneously performed both on a position operator and a momentum operator. And when one tried to measure the conjugated quantities sequently, the wavefunction of system will collapse strongly in the process of measurement. That is the reason why the CCR haven't been observed directly on experiments by far.

In the article, we try to measure the CCR within the correlation function of weak values in weak measurement, which called as the weak correlation. Weak measurement could fetch quantum information without collapse the measured states. Our setup is then presented in Fig.1a, and we select the states of the system three times, the initial selected state denoted as $|i\rangle$, the mid-selected intermediate state as $|f\rangle$ and finally the post-selected state denoted again as  $|i\rangle$. We will refer to the three strongly projected states as the pre-selected, mid-selected and post-selected quantum states, where the pre-selection and post-selection are projected to the same prepared state $|i\rangle$. Compare to the traditional weak measurement\cite{Yakir1988}, we denote the post-selection twice, that is, one projects strongly on $|f\rangle$ and the following back on $|i\rangle$. Now let us define the weak correlation function. With respect to these selected states, in the first half of our setup the weak value of the single weakly measured observable $\hat{\mathcal{O}}$ is defined as
\begin{equation}
\langle\hat{\mathcal{O}}\rangle_{w}=\frac{\langle f|\hat{\mathcal{O}}|i\rangle}{\langle f|i\rangle}=\mathcal{O}_w,
\end{equation}
while in the second half,
\begin{equation}
\langle\hat{\mathcal{O}}\rangle_{\bar{w}}=\frac{\langle i|\hat{\mathcal{O}}|f\rangle}{\langle i|f\rangle}=\mathcal{O}_{\bar{w}}.
\end{equation}
with the relation $\mathcal{O}_{\bar{w}}=\mathcal{O}_w^*$.\cite{Yakir2008}Then we define weak correlation function between the observables $\hat{A}$ and $\hat{B}$, which is given by
\begin{equation}
\langle\hat{A}\hat{B}\rangle_{w}=\frac{\langle i|\hat{A}|f\rangle\langle f|\hat{B}|i\rangle}{\langle i|f\rangle\langle f|i\rangle}=\langle A_{\bar{w}}B_w\rangle,
\end{equation}
where the observable $\hat{B}$ is weakly measured in the first half and while $\hat{A}$ is weakly measured in the second half of the experiment setup. Similarly, the weak correlation function between the observables $\hat{B}$ and $\hat{A}$ is given by
\begin{equation}
\langle\hat{B}\hat{A}\rangle_{w}=\frac{\langle i|\hat{B}|f\rangle\langle f|\hat{A}|i\rangle}{\langle i|f\rangle\langle f|i\rangle}=\langle B_{\bar{w}}A_w\rangle.
\end{equation}
The above two weak correlations between the observables $\hat{A}$ and $\hat{B}$ have the opposite order of measurement. Here we noted a recant work on the product of weak values.\cite{Hall2016} The high-order weak correlation functions are defined in Appendix A. Now the weak correlation of a commutator has the form
\begin{equation}
\langle[\hat{A},\hat{B}]\rangle_{w}=\langle\hat{A}\hat{B}-\hat{B}\hat{A}\rangle_{w}=\langle A_{\bar{w}}B_w\rangle-\langle B_{\bar{w}}A_w\rangle,
\end{equation}
and the relation with the normal expectation on pre-selected state is $\langle[\hat{A},\hat{B}]\rangle=\langle i|[\hat{A},\hat{B}]|i\rangle=\sum_{f}|\langle f|i\rangle|^2\langle[\hat{A},\hat{B}]\rangle_{w}$. 
If the commutator $[\hat{A},\hat{B}]=c$ is a number, then one expect 
\begin{equation}
\sum_{f}|\langle f|i\rangle|^2 (\langle[\hat{A},\hat{B}]\rangle_{w}-c)=0.
\end{equation}
With properly selected initial and finial state and satisfying the certain condition $|\langle f|i\rangle|^2\neq 0$, we could get the weak value of commutator. Applying the above formula to the canonical commutator $[\hat{x},\hat{p}]=i\hbar$, we obtain
\begin{equation}
i\hbar=\langle[\hat{x},\hat{p}]\rangle_{w}=\langle x_{\bar{w}}p_w\rangle-\langle p_{\bar{w}}x_w\rangle.
\end{equation}
\begin{figure}[t]
\begin{center}\includegraphics[width=.8\columnwidth]{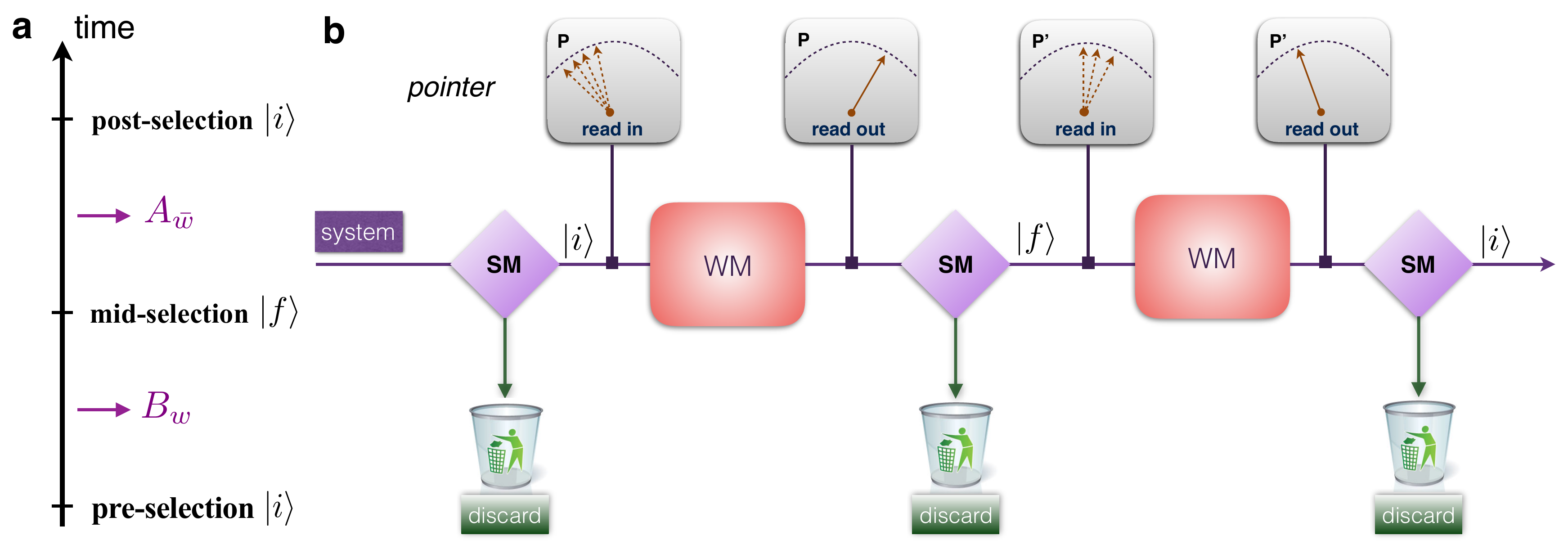}
\end{center}
\caption{Pre-mid-post-selection weak measurement setup to measure the weak correlations. (a) The weak measuring procedure with the pre-selection, mid-selection and post-selection. The weak value of operator $\hat{B}$ is weakly measured in the first half while that of operator $\hat{A}$ is weakly measured in the second half. Then the weak correlation $\langle\hat{A}\hat{B}\rangle_{w}$ is observed with properly selected on the desired wavefunctions. (b) The quantum system coupling weakly with the probe devices will read in/out a pointer without collapse the system state (the pointer $\mathbf{P}$ and $\mathbf{P'}$). Then the prepared wavefunction strongly project on the desired states, if three selections are all passed through, the pointers are kept, otherwise discarded. The two weakly coupling processes are the weak measurement (WM) and the three selections are the strong measurement (SM). The measurement setup could be extended to measure the higher-order weak correlations (see Appendix A).}
\label{fig:fig1}
\end{figure}
In general the weak value quantity is a complex number, so we attempt to rewrite the weak values with separated real part ($\Re$) and imaginary part ($\Im$),
\begin{subequations}
\begin{align}
      x_w&=\Re\{x_w\}+i\Im \{x_w\},\quad p_w=\Re\{p_w\}+i\Im \{p_w\},\\
      x_{\bar{w}}&=\Re\{x_w\}-i\Im \{x_w\},\quad p_{\bar{w}}=\Re\{p_w\}-i\Im \{p_w\},
\end{align}
\end{subequations}
where we apply the relations $x_{\bar{w}}=x_w^*, p_{\bar{w}}=p_w^*$. Then taking the imaginary part of equation (6), the weak correlation of the canonical commutator is then simplified as
\begin{equation}
\langle \Re\{x_w\}\Im \{p_w\}\rangle-\langle \Im \{x_w\}\Re\{p_w\}\rangle=\hbar/2.
\end{equation}
This is the final result of our setup to measure the fundamental canonical commutator in quantum mechanics with weak correlation of the position $\hat{x}$ and its conjugate momentum $\hat{p}$. For simplicity, suppose we prepare the proper selected states $|i\rangle$ and $|f\rangle$ to satisfy the condition $p_w=p_{\bar{w}}$, which implies $p_w$ real and its imaginary part $\Im \{p_w\}=0$. Then our final result could be further simplified as
\begin{equation}
\langle \Im \{x_w\}p_w\rangle=-\hbar/2.
\end{equation}
Now let us do the experiment as shown in Fig.1b. Suppose we want to weakly measure the position $\hat{x}$ of a system in the first half. We shall use an inaccurate pointer $\mathbf{P}$ as a measuring probe apparatus. Consider\cite{Yakir2008-2,Neumann1955}
\begin{equation}
H_{int}=-g(t) \hat{x}\otimes \hat{x}_d,
\end{equation}
where $\hat{x}_d$ is the position coordinate of the probe pointer, $g(t)$ is a coupling impulse function satisfying $\int_0^{T} g(t) dt=1$, and $T$ is the coupling time. We shall start the measuring process with the initial vector state $|i\rangle\otimes |\phi(x_d)\rangle$ where $\phi(x_d)$ is the initial state of the measuring device. Then we apply the interaction Hamiltonian and obtain
\begin{equation*}
e^{-iH_{int}t/\hbar}|i\rangle\otimes |\phi(x_d)\rangle.
\end{equation*}
Next, we interselect the state $|f\rangle$ by strongly measurement as shown in Fig.1b. Then after this one performs a projective measurement and get the outcome $|f\rangle\langle f|$, and the final state of pointer is given by
\begin{eqnarray}
&&|f\rangle\langle f|e^{-i\int H_{int}dt/\hbar}|i\rangle\otimes |\phi(x_d)\rangle \nonumber\\
&\approx&|f\rangle\langle f|(1+i\hat{x}\otimes \hat{x}_d/\hbar)|i\rangle\otimes |\phi(x_d)\rangle \nonumber\\
&=&|f\rangle\otimes\langle f|i\rangle(1+ix_w \hat{x}_d/\hbar) |\phi(x_d)\rangle \nonumber\\
&\approx&|f\rangle\otimes\left(\langle f|i\rangle e^{ix_w \hat{x}_d/\hbar}\right) |\phi(x_d)\rangle,
\end{eqnarray}
where $x_w=\frac{\langle f|\hat{x}|i\rangle}{\langle f|i\rangle}$. Take the initial state $|\phi(x_d)\rangle$ of $\mathbf{P}$ to have a Gaussian distribution
\begin{equation}
|\phi(x_d)\rangle=\frac{1}{(2\pi\sigma^2)^{1/4}}\exp\Big\{-\frac{x_d^2}{4\sigma^2}\Big\},
\end{equation}
with the width $\sigma$ of its probability distribution. Since the measurement is weak, there is a great uncertainty $\sigma$ in pointer position of $|\phi(x_d)\rangle$ when read-in. After measuring in the first half, the final state of pointer $\mathbf{P}$ will turn out to be
\begin{equation}
e^{ix_w \hat{x}_d/\hbar} |\phi(x_d)\rangle=\frac{1}{(2\pi\sigma^2)^{1/4}}\exp\Big\{-\frac{\left(x_d+2\sigma^2\Im\{x_w\}\right)^2}{4\sigma^2}\Big\}e^{ix_d\Re\{x_w\}},
\end{equation}
where the final expression could be found in the ref.\cite{Yakir1990}. This present that if we start with a Gassian pointer, its center will be shifted in real coordinate and phase coordinate as
\begin{subequations}
\begin{align}
 \Delta x_d&=-2\sigma^2\Im\{x_w\},\\
 \Delta {p_d}&=\Re\{x_w\},
\end{align}
\end{subequations}
where if in optical system we could apply a Fourier lens to image the position shift $\Delta{x_d}$ or the phase shift $\Delta{p_d}$ onto the CCD.\cite{Dressel2014,Steinberg2013}Therefore we could read out the imaginary part of $x_w$ from the probe pointer $\mathbf{P}$ directly with $\Im\{x_w\}=-\Delta{x_d}/2\sigma^2$.\\

Let us move forward to the second half, suppose we want to weakly measure the momentum $\hat{p}$ of the particle. Consider
\begin{equation}
H'_{int}=g'(t) \hat{p}\otimes \hat{p}'_d,
\end{equation}
where $g'(t)$ is the normalized coupling.\cite{Yakir2008-2}  And we use the above couping probe pointer $\mathbf{P'}$ as a measuring device and $\hat{p}_d$ is its momentum operator. Now we should begin with the measuring process with the intermediate state $|f\rangle\otimes |\phi'(x_d)\rangle$. Then we apply the interaction Hamiltonian to the intermediate state $e^{-iH'_{int}t/\hbar}|f\rangle\otimes |\phi'(x_d)\rangle$ and post-selected $|i\rangle$ again by strongly measured, then
\begin{eqnarray}
&&|i\rangle\langle i|e^{-i\int H'_{int}dt/\hbar}|f\rangle\otimes |\phi'(x_d)\rangle \nonumber\\
&\approx&|i\rangle\langle i|(1-i\hat{p}\otimes \hat{p}'_d/\hbar)|f\rangle\otimes |\phi'(x_d)\rangle \nonumber\\
&=&|i\rangle\otimes\langle i|f\rangle(1-ip_{\bar{w}} \hat{p}'_d/\hbar) |\phi'(x_d)\rangle \nonumber\\
&\approx&|i\rangle\otimes\langle i|f\rangle e^{-ip_{\bar{w}} \hat{p}'_d/\hbar} |\phi'(x_d)\rangle\nonumber\\
&=&|i\rangle\otimes\langle i|f\rangle  |\phi'(x_d-p_{\bar{w}})\rangle,
\end{eqnarray}
with $p_{\bar{w}}=\frac{\langle i|\hat{p}|f\rangle}{\langle i|f\rangle}$ and the initial state of pointer $\mathbf{P'}$ is $\phi'(x_d)=\frac{1}{(2\pi\sigma'^2)^{1/4}}\exp\big\{-\frac{x_d^2}{4\sigma'^2}\big\}$ with width $\sigma'$. If choosing the proper initial and finial states $|i,f\rangle$ to obtain $p_w=p_{\bar{w}}$, then one could read out the weak value $p_w=\Delta{x'_d}$ from the probe pointer $\mathbf{P'}$ with the coordinate shift $\Delta'{x_d}$ directly. To conclude the experiment, we predict that
\begin{equation}
\langle\Delta{x_d}\Delta x'_d\rangle=\hbar \sigma^2,
\end{equation}
where the correlation function of the measured $\mathbf{P}$-shift and $\mathbf{P'}$-shift equals to the product of reduced Planck constant and initial Gaussian variance $\sigma^2$ of the first pointer $\mathbf{P}$. This experiment could be performed hopfully to test the cornerstone of quantum mechanics in the near future.

\section{Measuring the Riemann operator $\mathcal{\hat{R}}=\frac{1}{2}+i\hat{\rho}$}
We define the Riemann operator
\begin{equation}
\mathcal{\hat{R}}=\frac{i\hat{p}\hat{x}}{\hbar},
\end{equation}
has a set of eigenstates $|\phi_n\rangle$
\begin{equation*}
\mathcal{\hat{R}}|\phi_n\rangle=\mathcal{R}_n|\phi_n\rangle,
\end{equation*}
where the eigenvalue spectrum of $\mathcal{\hat{R}}$ are the nontrivial zeros of Riemann zeta function $\zeta (s)$ in the framework of the Hilbert-Polya conjecture.\cite{Edwards1974, Montgomery1973} The Riemann hypothesis (RH) states that the nontrivial complex zeros of $\zeta (s)$ on the critical line $\Re\{s\}=\frac{1}{2}$; that is, nonimaginary solution $\rho_n$ of
\begin{equation}
\zeta (\mathcal{R}_n)=\zeta (1/2+i\rho_n)=0,
\end{equation}
are all real and for instances the first three zeros $\rho_1=14.13, \rho_2=21.02, \rho_3=25.01 \cdots$ and so on. 

Within our assumption, we could prove that the canonical commutation relation $[\hat{x},\hat{p}]=i\hbar$ in quantum mechanics is intimately equivalent to the RH. Firstly, we assume that the RH holds true, then the Riemann operator has eigenvalues
\begin{equation}
\mathcal{\hat{R}}|\phi_n\rangle=\left(\frac{1}{2}+i\rho_n\right)|\phi_n\rangle,
\end{equation}
For arbitrary quantum state $|\psi\rangle=\sum_n c_n|\phi_n\rangle$, we obtain
\begin{equation*}
\mathcal{\hat{R}}|\psi\rangle=\frac{1}{2}|\psi\rangle+i\sum_n c_n \rho_n|\phi_n\rangle,
\end{equation*}
and for the complex-conjugated operator $\mathcal{\hat{R}}^{\dagger}=-i\hat{x}\hat{p}/\hbar$
\begin{equation*}
\mathcal{\hat{R}}^{\dagger}|\psi\rangle=\frac{1}{2}|\psi\rangle-i\sum_n c_n \rho_n|\phi_n\rangle.
\end{equation*}
Then add them together we find that
\begin{equation}
\left(\frac{\mathcal{\hat{R}}+\mathcal{\hat{R}}^{\dagger}}{2}-\frac{1}{2}\right)|\psi\rangle=0.
\end{equation}
Due to $|\psi\rangle$ arbitrary, then $\left(\frac{\mathcal{\hat{R}}+\mathcal{\hat{R}}^{\dagger}}{2}-\frac{1}{2}\right)=0$. Thus one could derive the canonical commutation relation $[\hat{x},\hat{p}]=i\hbar$. Secondly, we assume that the canonical commutation relation holds true, then the Riemann operator could be expressed as
\begin{equation}
\mathcal{\hat{R}}=\frac{i\hat{p}\hat{x}}{\hbar}=\frac{i\left(-[\hat{x},\hat{p}]+\{\hat{x},\hat{p}\}\right)}{2\hbar}=\frac{1}{2}+i\frac{\{\hat{x},\hat{p}\}}{2\hbar}=\frac{1}{2}+i\hat{\rho},
\end{equation}
where the anticommutator $\hat{\rho}=\frac{\{\hat{x},\hat{p}\}}{2\hbar}$ is Hermitian and its discrete eigenvalues are all real with properly quantized.\cite{Berry1990} Therefore, for eigenstate $|\phi_n\rangle$, we obtain that $\mathcal{\hat{R}}|\phi_n\rangle=\left(\frac{1}{2}+i\hat{\rho}\right)|\phi_n\rangle=\left(\frac{1}{2}+i\rho_n\right)|\phi_n\rangle$ where $\rho_n$ are all real. Thus the imaginary part of discrete eigenstates of $\mathcal{\hat{R}}$ through the Berry-Keating conjecture\cite{Berry1990} , that is, the nontrivial complex zeros of $\zeta(s)=0$, lie on the critical line $\Re\{\mathcal{\hat{R}}\}=1/2$, which is the same statement as the RH.

Unfortunately, how to quantize $\hat{\rho}$ is still an open problem.\cite{Berry1990, Sierra2008}  And our argument is based currently on the Berry-Keating conjecture which have been proved yet. In general, the Riemann zero operator $\mathcal{\hat{R}}$ would be connected with the nontrivial zeros of general Riemann hypothesis of Dirichlet $\mathbf{ L}$-function with an unknown character $\chi$ which demands more works(see Appendix B). 

The equivalence demonstrates here, if properly quantized $\mathcal{\hat{R}}$ with eigenvalues lies exactly on the nontrivial zeros of zeta function, then one could claim that Riemann hypothesis will been proven quantum mechanically. Conversely if RH is wrong, then there are a couple of zeros out of critical line (due to the Riemann $\xi$-function $\xi(1-s)=\xi(s)$) and the corresponding eigenstate satisfies $\mathcal{\hat{R}}|\phi_n'\rangle=\left(\frac{1}{2}\pm\delta+i\rho_n'\right)|\phi_n'\rangle$ where $\delta\neq 0$. Thus at the quantum state of the special zeros, the expectation $\langle\phi_n'|[\hat{x},\hat{p}]|\phi_n'\rangle=i\hbar (1\pm 2\delta)\neq i\hbar$ shows the postulate of quantum mechanics is not always testing-robust for arbitrary quantum state, in other word, it denies quantum mechanics.

In our framework of weak measurement, we could observe the weak value of $\hat{\rho}$
\begin{equation}
\rho_w=\left\langle\frac{\{\hat{x},\hat{p}\}}{2\hbar}\right\rangle_{w}=\frac{\langle x_{\bar{w}}p_w\rangle+\langle p_{\bar{w}}x_w\rangle}{2\hbar}.
\end{equation}
Performing the experiment as we shown in fig. 1b, one gets
\begin{equation}
\langle \Re\{x_w\}\Re \{p_w\}\rangle+\langle \Im \{x_w\}\Im \{p_w\}\rangle=\hbar \rho_w.
\end{equation}
where $\rho_w=\frac{\langle f| \hat{\rho}|i\rangle}{\langle f|i\rangle}=\frac{\sum_n \rho_nc_{n}^{(f)*}c_{n}^{(i)}}{\sum_n c_{n}^{(f)*}c_{n}^{(i)}}$ and $|i,f\rangle=\sum_n c_{n}^{(i,f)}|\phi_n\rangle$. In our setup one could observe $\hbar \rho_w=\langle\Delta{p_d}\Delta x'_d\rangle$, and expect that the weak measurement of $\rho_w$ would shed new light on the spectral realization of nontrivial Riemann zeros. However we have no idea how to reconstruct the spectrum ($\rho_n$) from the weak value $\rho_w$ but still could fetch some information when properly selected states. In short, our main result as we presented allows us to observe the weak value of Riemann operator $\mathcal{R}_w=1/2+i\rho_w$, which yields experimentally to the fundamental physical problems of measurability of the CCR postulate of quantum mechanics (see eq. 9) and the Riemann hypothesis of complex analysis and number theory (see eq. 25).

\section{Measuring Pauli operators $[\sigma_x, \sigma_y]=2i\sigma_z$}
Applying our definition of weak correlation to the Stern-Gerlach experiment, we could check the commutator and anti-commutator of Pauli operators in spin half systems, where firstly write down all Pauli operators
\begin{equation}
\sigma_x=\left(\begin{array}{cc}0 & 1 \\1 & 0\end{array}\right), \sigma_y=\left(\begin{array}{cc}0 & -i \\i & 0\end{array}\right), \sigma_z=\left(\begin{array}{cc}1 & 0 \\0 & -1\end{array}\right).
\end{equation}
Now we are about to measure the weak correlation of the anticommutator and commutator,
\begin{subequations}
\begin{align}
&\langle\{\sigma_x, \sigma_y\}\rangle_w=0,\\
 &\langle[\sigma_x, \sigma_y]\rangle_w=2i\langle\sigma_z\rangle_w.
\end{align}
\end{subequations}
Following the landmark paper of Aharonov, et al., we choose the selected states\cite{Yakir1988}
\begin{equation}
|i\rangle=\frac{1}{\sqrt{2}}\left(\begin{array}{c}\cos \alpha/2+\sin \alpha/2 \\\cos \alpha/2-\sin \alpha/2\end{array}\right),\quad |f\rangle=\frac{1}{\sqrt{2}}\left(\begin{array}{c}1 \\1\end{array}\right),
\end{equation}
where the initial spin pointed in the xz plane has an angle $\alpha$ with axis $x$. Now we design the experiment setup with the pre-selection $|i\rangle$, mid-selection $|f\rangle$ and post-selection $|i\rangle$. Performing the experiment, one could measure
\begin{subequations}
\begin{align}
\langle\sigma_x\sigma_y\rangle_w&=\frac{\langle i|\sigma_x|f\rangle\langle f|\sigma_y|i\rangle}{\langle i|f\rangle\langle f|i\rangle}=i\tan \alpha/2,\\
\langle\sigma_y\sigma_x\rangle_w&=\frac{\langle i|\sigma_y|f\rangle\langle f|\sigma_x|i\rangle}{\langle i|f\rangle\langle f|i\rangle}=-i\tan \alpha/2,
\end{align}
\end{subequations}
and then independently
\begin{equation}
\langle\sigma_z\rangle_w=\frac{\langle f|\sigma_z|i\rangle}{\langle f|i\rangle}=\tan \alpha/2.
\end{equation}
Thus we obtain the weak correlation functions
\begin{subequations}
\begin{align}
\langle\{\sigma_x, \sigma_y\}\rangle_w&=i\tan \alpha/2-i\tan \alpha/2=0,\\
 \langle[\sigma_x, \sigma_y]\rangle_w&=2i\tan \alpha/2=2i\langle\sigma_z\rangle_w.
\end{align}
\end{subequations}
Therefore we pass successfully the testing of the anticommutator and commutator of Pauli operators in our weak measurement setup, where the weak value and weak correlations inherit the algebra of Pauli operators. However, it should be noted that in general the (anti-)commutation relations may be failed to establish in the experiment setup. The reason is beacuse the weak values and weak correlations are closely dependent on the proper selections of quantum systems and the selection is crucial to the weak measurement. And more discussions about weak correlations see the following Appendix A.

\newpage
\appendix
\section{The simultaneity of weak measurement and high-order weak correlation functions}
Suppose we pre-select the initial state $|i\rangle$ at time $t_0$, mid-select the following final state $|f\rangle$ at time $t_1$ and then post-select the initial state $|i\rangle$ at the last stage. Then we weakly measure the operator $\hat{B}(t)$ at the first half when $t_0<t<t_1$ and then weakly measure the operator $\hat{A}(t')$ at the second half when $t_1<t'<t_2$.  Now we are going to proof the simultaneity of weak correlation function $\langle\hat{A}(t')\hat{B}(t)\rangle_{w}=\langle\hat{A}\hat{B}\rangle_{w}$. Consider the definition of weak correlation
\begin{eqnarray}
\langle\hat{A}(t')\hat{B}(t)\rangle_{w}&=&\frac{\langle i|\hat{A}(t')|f\rangle\langle f|\hat{B}(t)|i\rangle}{\langle i|f\rangle\langle f|i\rangle}\nonumber\\
&=&\left(\lim_{t'\rightarrow t_1^+}\frac{\langle i|\hat{A}(t')|f\rangle}{\langle i|f\rangle}\right)\left(\lim_{t\rightarrow t_1^-}\frac{\langle f|\hat{B}(t)|i\rangle}{\langle f|i\rangle}\right)\nonumber\\
&=&\frac{\langle i|\hat{A}(t_1^+)|f\rangle\langle f|\hat{B}( t_1^-)|i\rangle}{\langle i|f\rangle\langle f|i\rangle}\nonumber\\
&=&\langle\hat{A}(t_1^+)\hat{B}(t_1^-)\rangle_{w}\nonumber\\
&=&\langle\hat{A}\hat{B}\rangle_{w}
\end{eqnarray}
where $t_1$ is the time of mid-selection but still arbitrary between the preparation and post-selection. Therefore, the weak correlation between the operators $\hat{A}$ and $\hat{B}$ is simultaneously observable but physically measured in order.\\

By performing the selection processes as follows $|i\rangle, |f\rangle, |i\rangle, |f\rangle,|i\rangle, \cdots$, as shown the experiment timeline in Fig.2, we could define the high-order weak correlation functions, correspondingly
\begin{subequations}
\begin{align}
      \langle\hat{\mathcal{O}}^{(2N)}\hat{\mathcal{O}}^{(2N-1)}\cdots\hat{\mathcal{O}}^{(2)}\hat{\mathcal{O}}^{(1)}\rangle_{w}&=\langle\hat{\mathcal{O}}_{\bar{w}}^{(2N)}\hat{\mathcal{O}}_w^{(2N-1)}\cdots\hat{\mathcal{O}}_{\bar{w}}^{(2)}\hat{\mathcal{O}}_w^{(1)}\rangle,\\
      \langle\hat{\mathcal{O}}^{(2N+1)}\hat{\mathcal{O}}^{(2N)}\cdots\hat{\mathcal{O}}^{(2)}\hat{\mathcal{O}}^{(1)}\rangle_{w}&=\langle\hat{\mathcal{O}}_w^{(2N+1)}\hat{\mathcal{O}}_{\bar{w}}^{(2N)}\cdots\hat{\mathcal{O}}_{\bar{w}}^{(2)}\hat{\mathcal{O}}_w^{(1)}\rangle.
\end{align}
\end{subequations}
where $N$ is a positive integer. The definition of high-order weak correlation still satisfies the simultaneity, the proof is similar to what we did for the second-order weak correlation. The high-order weak correlation (include the 2-order weak correlation) could be expected helpfully to explore the quantum entanglements and correlations in quantum systems.\\
\begin{figure}[t]
\begin{center}\includegraphics[width=.43\columnwidth]{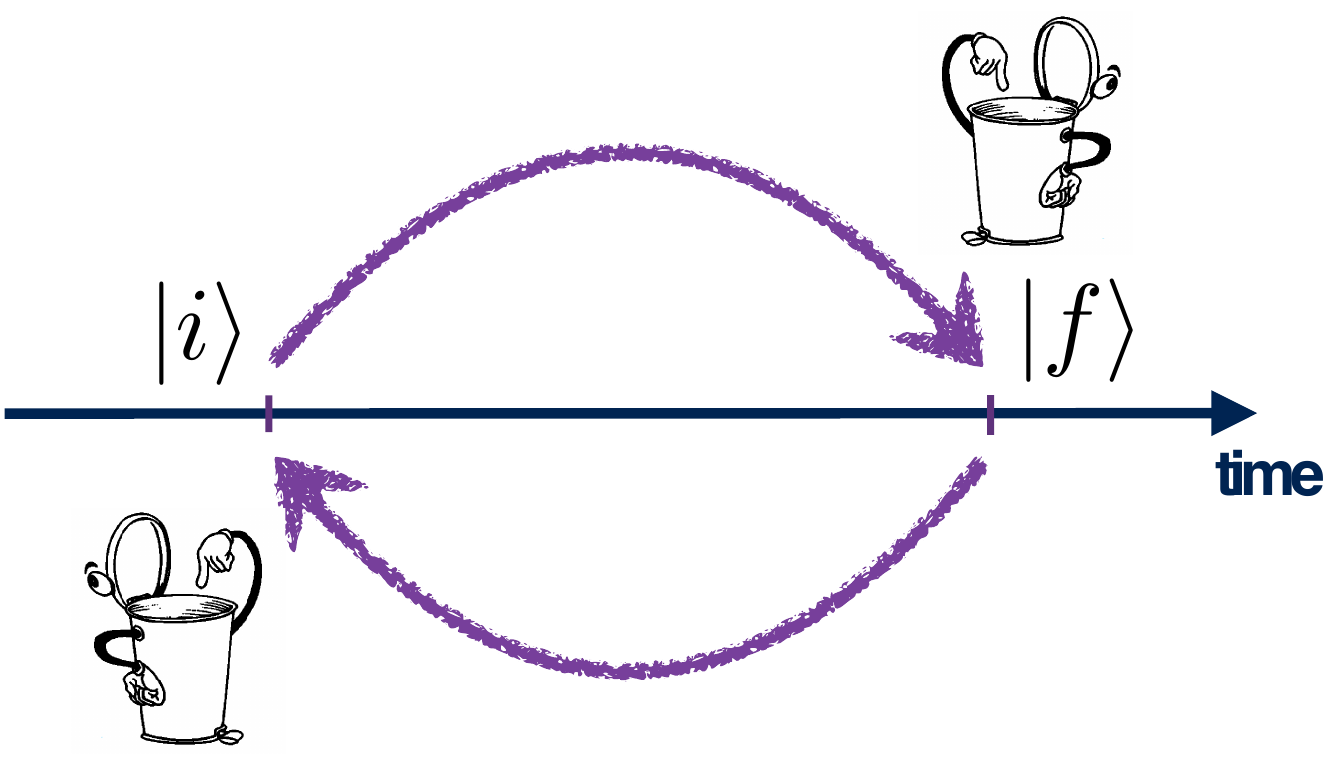}
\end{center}
\caption{Timeline of pre-mid-post-selection weak measurement to perform high-order weak correlation functions in further experiments.}
\label{fig:fig2}
\end{figure}

To see some basic properties of weak correlations, let us define a dual procedure as pre-selection $|f\rangle$, mid-selection $|i\rangle$ and post-selection $|f\rangle$ and so on, which interchanges the selected states $|i\rangle$ and $|f\rangle$.\cite{Yakir2008}Then we will measure the correlation function of weak values of $\hat{A}$ and $\hat{B}$ on this interchanged procedure. The dual weak correlation is given by
\begin{equation}
\langle\hat{B}\hat{A}\rangle_{\bar{w}}=\frac{\langle f|\hat{B}|i\rangle\langle i|\hat{A}|f\rangle}{\langle f|i\rangle\langle i|f\rangle}=\langle B_{w}A_{\bar{w}}\rangle=\langle A_{\bar{w}}B_{w}\rangle=\langle\hat{A}\hat{B}\rangle_{w}.
\end{equation}
In general, for the high-order weak correlations, the symmetries are given by
\begin{subequations}
\begin{align}
      \langle\hat{\mathcal{O}}^{(2N)}\hat{\mathcal{O}}^{(2N-1)}\cdots\hat{\mathcal{O}}^{(2)}\hat{\mathcal{O}}^{(1)}\rangle_{w}&=\langle\hat{\mathcal{O}}^{(1)}\hat{\mathcal{O}}^{(2)}\cdots\hat{\mathcal{O}}^{(2N-1)}\hat{\mathcal{O}}^{(2N)}\rangle_{\bar{w}},\\
     \langle\hat{\mathcal{O}}^{(2N+1)}\hat{\mathcal{O}}^{(2N)}\cdots\hat{\mathcal{O}}^{(2)}\hat{\mathcal{O}}^{(1)}\rangle_{w}&=\langle\hat{\mathcal{O}}^{(2N+1)}\hat{\mathcal{O}}^{(2N)}\cdots\hat{\mathcal{O}}^{(2)}\hat{\mathcal{O}}^{(1)}\rangle_{w},\\
      \langle\hat{\mathcal{O}}^{(2N+1)}\hat{\mathcal{O}}^{(2N)}\cdots\hat{\mathcal{O}}^{(2)}\hat{\mathcal{O}}^{(1)}\rangle_{\bar{w}}&=\langle\hat{\mathcal{O}}^{(2N+1)}\hat{\mathcal{O}}^{(2N)}\cdots\hat{\mathcal{O}}^{(2)}\hat{\mathcal{O}}^{(1)}\rangle_{\bar{w}}.
\end{align}
\end{subequations}
and finally for the weak correlation of commutators, one finds that the identity
\begin{equation}
\langle[\hat{A},\hat{B}]\rangle_{w}=-\langle[\hat{A},\hat{B}]\rangle_{\bar{w}},
\end{equation}

\newpage
\section{The generalized Riemann hypothesis and the Dirichlet $\mathcal{L}$-functions}
The Riemann hypothesis is one of the most important conjectures in mathematics. It is a statement about the zeros of the Riemann zeta function lie on the critical line. Various geometrical and arithmetical objects can be described by so-called global $\mathcal{L}$-functions, which are formally similar to the Riemann zeta-function. 

When the Riemann hypothesis is formulated for Dirichlet $\mathcal{L}$-functions, it is known as the generalized Riemann hypothesis (GRH).\cite{Edwards1974, Wikipedia} The formal statement of the GRH follows: A Dirichlet character is a completely multiplicative arithmetic function $\chi$ such that there exists a positive integer k with $\chi(n + k) = \chi(n)$ for all n and $\chi(n) = 0$ whenever $\gcd(n, k) > 1$. If such a character is given, we define the corresponding Dirichlet $\mathcal{L}$-function by
\begin{equation}
\mathcal{L}(\chi,s)=\sum_{n=1}^{\infty}\frac{\chi(n)}{n^s}
\end{equation}
for every complex number s with $\Re\{s\}>1$. By analytic continuation, this function can be extended to a meromorphic function defined on the whole complex plane except s=1. One can then ask the same question about the zeros of these $\mathcal{L}$-functions, yielding various generalizations of the hypothesis. The GRH asserts that for every Dirichlet character $\chi$ and every complex number s with $\mathcal{L}(\chi,s)=0$, if the real part of s is in the critical stripe between 0 and 1, then it is actually on the critical line $\Re\{s\}=1/2$. For example, the case $\chi(n) = 1$ yields the ordinary Riemann hypothesis.

\begin{acknowledgments}
We acknowledge Yakir Aharonov for useful discussions and comments. This work was supported by the PBC program of the Israel council of higher education.
\end{acknowledgments}

\end{document}